\documentclass{article}
\usepackage{graphicx}
\usepackage{verbatim}

\usepackage{amsmath,amssymb,amsfonts,amscd}
\usepackage{graphics}
\usepackage{natbib}
\usepackage{hyperref}

\hypersetup{colorlinks=true, citecolor=blue, linkcolor=blue, urlcolor=blue, filecolor=black} 

\begin{document}

\title{Homophily and Contagion Are Generically Confounded in Observational Social Network Studies}
\author{Cosma Rohilla Shalizi\\ {\small Department of Statistics, Carnegie Mellon University}\\
{\small Santa Fe Institute}
\and
Andrew C. Thomas\\
{\small Department of Statistics, Carnegie Mellon University}}
\date{Draft begun 13 November 2009, last \LaTeX 'd \today}
\maketitle

\begin{abstract}
  We consider processes on social networks that can potentially involve three
  factors: homophily, or the formation of social ties due to matching
  individual traits; social contagion, also known as social influence; and the
  causal effect of an individual's covariates on their behavior or other
  measurable responses. We show that, generically, all of these are confounded
  with each other.  Distinguishing them from one another requires strong
  assumptions on the parametrization of the social process or on the adequacy
  of the covariates used (or both).  In particular we demonstrate, with simple
  examples, that asymmetries in regression coefficients cannot identify causal
  effects, and that very simple models of imitation (a form of social
  contagion) can produce substantial correlations between an individual's
  enduring traits and their choices, even when there is no intrinsic affinity
  between them.  We also suggest some possible constructive responses to these
  results.
\end{abstract}

\tableofcontents

\section{Introduction: ``If your friend jumped off a bridge, would you jump too?''}

We all know that people who are close to each other in a social network are
similar in many ways: they share characteristics, act in similar ways, and
similar events are known to befall them.  Do they act similarly because they
are close in the network, due to some form of influence that acts along network
ties (or, as it is often suggestively put, ``contagion''\footnote{Analogies
  between the spread of ideas and behaviors --- especially {\em disliked} ideas
  and behaviors --- and the spread of disease are ancient.  Pliny the Younger,
  for instance, referred to Christianity as a ``contagious superstition'' in a
  letter to the Emperor Trajan in 110 ({\em Epistles} X 96).
  \citet{Siegfried-germs-and-ideas} gives further examples.  The best treatment
  of this analogy is made by \citet{Sperber-explaining}.})?  Or rather are they
close in the network {\em because} of these similarities, through the processes
known assortative mixing on traits, or more simply as homophily
\citep{Birds-of-a-Feather-review}?  Suppose that there are two friends named
Ian and Joey, and Ian's parents ask him the classic hypothetical of social
influence: ``If your friend Joey jumped off a bridge, would you jump too?'' Why 
might Ian answer ``yes''?
\begin{enumerate}
\item Because Joey's example inspired Ian (social contagion/influence);
\item Because Joey infected Ian with a parasite which suppresses fear of
  falling (biological contagion);
\item Because Joey and Ian are friends on account of their shared fondness for
  jumping off bridges (manifest homophily, on the characteristic of interest);
\item Because Joey and Ian became friends through a thrill-seeking club, whose
  membership rolls are publicly available (secondary homophily, on a different
  yet observed characteristic);
\item Because Joey and Ian became friends through their shared fondness for
  roller-coasters, which was caused by their common thrill-seeking propensity,
  which also leads them to jump off bridges (latent homophily, on an unobserved 
  characteristic); 
\item Because Joey and Ian both happen to be on the Tacoma Narrows Bridge in
  November, 1940, and jumping is safer than staying on a bridge that is tearing
  itself apart (common external causation).
\end{enumerate}

The distinctions between these mechanisms --- and others which no doubt occur
to the reader --- are all ones which make {\em causal} differences.  In
particular, if there is any sort of contagion, then measures which specifically
prevent Joey from jumping off the bridge (such as restraining him) will also
have the effect of tending to keep Ian from doing so; this is not the case if
contagion is absent.  However, the crucial question is whether these
distinctions make differences in the purely observational setting, since we are
usually not able to conduct an experiment in which we push Joey off the bridge
and see whether Ian jumps (let alone repeated trials.)

The goal of this paper is to establish that these are, by and large,
phenomena that are surprisingly difficult to distinguish in purely 
observational studies.  More precisely, latent homophily
and contagion are generically confounded with each other (section
\ref{sec:homophily-fakes-contagion}), and any direct contagion effects cannot
be nonparametrically identified from observational data\footnote{We remind the
  reader of the relevant sense of ``identification''
  \citep{Manski-identification-for-prediction}.  We have a collection of random
  variables, which are generated by one causal process $M$ out of a set of
  possible processes $\mathcal{M}$.  Not all aspects of this process are
  recorded, and the result is a distribution $P$ over observables.  Each $M$
  leads to only one distribution over observables, $P(M)$.  A functional
  $\theta$ of the data-generating process is identifiable if it depends on $M$
  only through $P(M)$, i.e., if $\theta(M) \neq \theta(M^{\prime})$ implies
  $P(M) \neq P(M^{\prime})$.  Otherwise, the functional is unidentifiable.  If
  $\theta$ is identifiable only when $\mathcal{M}$ is restricted to a finitely
  parameterized family, then $\theta$ is parametrically identifiable (within
  that family).  If $\theta$ is identifiable without such a restriction, it is
  non-parametrically identifiable.  See further \citet[ch.\
  3]{Pearl-causality-2nd} on identification of causal effects from
  observables.}.  To identify contagion effects, we need either strong
parametric assumptions or strong substantive knowledge that lets us rule out
latent homophily as a causal factor.  It has been proposed that asymmetries in
regression estimates which match asymmetries in the social network would let us
establish direct social contagion; we show (Section \ref{sec:asymmetry}) as a
corollary of our main result that this also fails.

We realize that many issues with unobservable characteristics exist in many
observational study settings, not just in those that share our explicit focus
on network phenomena, yet our investigations of social contagion are not driven
by some animus; we are just as concerned for those investigations that ignore
network structure when it is present.  If contagion works along with homophily,
we show that it confounds inferences for relationships between homophilous
traits and outcome variables such as observed behaviors (Section
\ref{sec:contagion-fakes-causation}).  In particular, even when the true causal
effect of the homophilous trait is zero, the trait can still act as a strong
predictor of the outcome of interest merely through the outcome's natural
diffusion in a network (Section \ref{sec:voter-model}).

We also realize that our main findings are negative, and implicitly critical of
much previous work.  Section \ref{sec:positives} suggests some possible
constructive responses to our findings, while Section \ref{sec:conclusion}
concludes with some methodological reflections.

\subsection{Notation, Terminology, Conventions}

In our framework the random variable $X_i$ is a collection of unchanging latent
traits for node $i$; similarly, $Z_i$ is a collection of static observed
traits.  Both $X$ and $Z$ may be discrete, continuous, mixtures of both, etc.
The social network is represented by the binary variable $A_{ij}$, which is 1
if there is a (directed) edge from $i$ to $j$ --- that is, $i$ considers $j$ to
be a ``friend'' --- and 0 otherwise.  Time $t$ advances in discrete steps of
equal duration; this is inessential but avoids mathematical complications.
$Y_i(t)$ denotes a response variable for node $i$ at time $t$; again, whether
categorical, metric or otherwise doesn't matter.  (We will sometimes write this
as $Y(i,t)$ or even $Y_{it}$, as typographically convenient, and likewise for
other indices.)  These variables are also listed in Figure
\ref{table:terminology}, alongside a graphical representation of the
prototypical process we are examining.

We conducted all simulations in \texttt{R} \citep{R}.  Our code is available
from \url{http://www.stat.cmu.edu/~cshalizi/homophily-confounding/}.

\section{How Homophily and Individual-Level Causation Look Like
  Contagion} 
\label{sec:homophily-fakes-contagion}

The members of a social network often exhibit correlated behavior.  When we
speak of contagion or influence within networks, we imply that conditioning on
all other factors, there will be a temporal relationship between the behaviour
of individual $i$ at time $t$ and any neighbours of $i$ (potential $j$'s) at
the previous time point. This is easiest to see when all other causes of
adoption of a trait aside from the network itself are eliminated, such as
person-to-person infectious diseases
\citep{Bartlett-stochastic-population-models, Ellner-Guckenheimer,
  MEJN-epidemics-on-networks}, though other examples include the spread of
innovations \citep{Rogers-diffusion-of-innovations}.

More puzzling are situations such as the investigation of
\citet{Christakis-Fowler-spread-of-obesity}, where the behavior that apparently
spreads through the network is ``becoming obese'', as obesity is not normally
thought of as an infectious condition\footnote{There are however claims in the
  medical literature \citep{Atkinson-viruses-obesity} that certain viruses
  induce obesity in rodents and may contribute to the condition in human
  beings.  (Thanks to Matthew Berryman and Gustavo Lacerda for bringing this to
  our attention.)  We lack the knowledge to assess the soundness of these
  claims, let alone their plausibility as explanations of human obesity.}, or
the apparent spread of ``happiness'', documented by
\citet{Fowler-Christakis-happiness}.  It is natural to ask how much of such
``network autocorrelation'' --- the tendency of these behaviors to be
correlated in individuals that are closely connected --- is due to some direct
influence of $i$'s neighbors on $i$'s behavior, as opposed to the effect of
homophily, in which social ties form between individuals with similar
antecedent characteristics, who may then behave similarly as a
result\footnote{\citet[ch.\ 5]{Sperber-explaining} is a detailed and subtle
  exploration of just how powerful the latter mechanism can be, and how it can
  interact with imitation or contagion.}.

Social network scholars have long been concerned with this issue, under the
label of ``selection versus influence'' or ``homophily versus contagion''
\citep{Leenders-structure-and-influence}, usually with regard to {\em manifest}
homophily but certainly not limited to it. To give just one example of a
sophisticated recent attempt to divide the credit for network autocorrelation
between homophily and contagion, consider \citet{aral2009distinguishing}.  (The
following remarks apply, with suitable changes, to many other high-quality
studies, e.g., \citet{Bakshy-Karrer-Adamic,flickr-tagging-influence,%
  Yang-Longini-Halloran-detecing-transmission,bramoulle2009ipetsn}.)  They
worked with a uniquely obtained data set with a clear outcome measure: the
adoption of an online service over time, with users of an instant messaging
service as the (extremely large) community of interest.  To separate the
effects of contagion from those of homophily, a large and rich table of
covariates on an individual's personal and network characteristics was
assembled (with 46 covariates in total), and matched pairs were assembled using
propensity score estimation \citep{Rosenbaum-Rubin-1983} so that one member of
the pair had, at one point, exposure to the online service through one (or
more) of their network neighbors; assuming that these characteristic
differences had then been teased out, the difference in the adoption rate would
then reflect the total proportion of the adoption by contagion, allowing for an
estimate of the proportion of association that is attributable to contagion, as
opposed to the proportions caused by homophily, either secondary (in terms of
the 46 observed network characteristics) or manifest (caused by two users
becoming friends specifically due to their connection on the online service) --
but notably, not latent homophily, which may still remain as a component of the
so-called ``contagious'' proportion; this is due to the nature of propensity
score matching, which can simplify the relationships between {\em observed}
properties and the adoption of a ``treatment'' (in this case, network-localized
exposure to the service), the effort may prove to be inadequate if any
unobserved covariates have a part in both tie selection and in service
adoption.

This brings us to our fundamental point: to attempt to assign strengths to
influence or contagion as opposed to homophily presupposes that the distinction
is identifiable, and there have been grounds to doubt this for some time.
\citet{Manski-reflection-problem}, in a well-known paper, considered the
related problem of the identification of {\em group} effects: supposing that an
individual's behavior depends on some individual-level predictors and on the
mean behavior of the group to which they belong, can the degree of dependence
on the group be identified?  He showed that in general the answer is ``no'',
unless you make strong parametric assumptions, and perhaps not even then (since
group effects can fail to be identified even in linear models). Indeed, this
has been shown to cause difficulties in other social situation where this sort
of phantom influence can be observed: among others,
\citet{Calvo-Armengol-Jackson-on-parent-child-correlation} note that estimating
the apparent effect of parental influence on their child's educational outcomes
is confounded by the actions of the larger community.  (See
\citealt{Blume-Brock-Durlauf-Ioannides-identification} for a recent review of
the group-effects literature.)  However, this does not quite answer our
questions, since Manski considered influence from the group average, rather
than from individual members of the network neighborhood, and one could hope
this would provide enough extra information for identification.

We now show that, in fact, contagion effects are nonparametrically
unidentifiable in the presence of latent homophily --- that there is just no
way to separate selection from influence observationally.  Our proof involves
some simple manipulations of graphical causal models; we refer the reader to
standard references \citep{Spirtes-Glymour-Scheines, Pearl-causality-2nd,
  Pearl-on-causal-inference, Morgan-Winship-counterfactuals} for the necessary
background.

\subsection{Contagion Effects are Nonparametrically Unidentifiable}
\label{sec:contagion-unidentifiable}

\begin{figure}
\begin{center}
\begin{tabular}{cc}
\begin{minipage}[b]{0.5\linewidth}\centering
\resizebox{\textwidth}{!}{\includegraphics{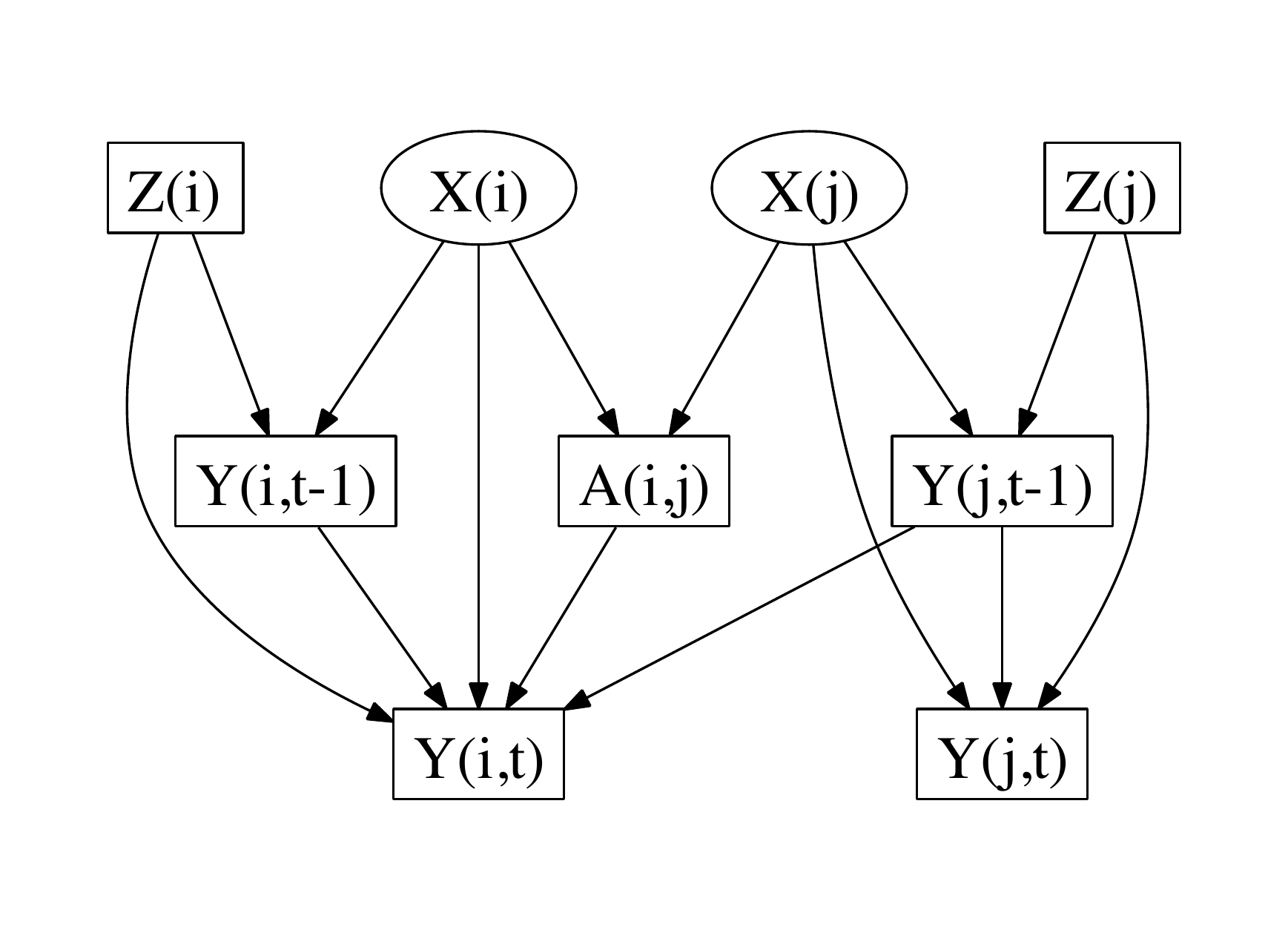}}
\caption{Causal graph allowing for latent variables ($X$) to influence both
  manifest network ties $A_{ij}$ and manifest behaviors ($Y$).}
\label{fig:latent-homophily}
\end{minipage}

&
\hspace{0.2cm}

\begin{minipage}[b]{0.42\linewidth}\centering
\begin{tabular}{cc}
Symbol & Meaning \\
\hline
$i,j$ & Individuals \\
$Z$ & Observed Traits \\
$X$ & Latent Traits \\
$Y$ & Observed Outcomes
\end{tabular}
\vspace{1.5cm}
\caption{Notational guide to terms used in this
  investigation.\label{table:terminology}}
\end{minipage}
\end{tabular}
\end{center}
\end{figure}

We first assume that there is latent homophily present in the system: the 
network tie $A_{ij}$ is influenced by the unobserved traits of each
individual, $X_i$ and $X_j$.  We assume that the ``past'' observable outcome
$Y_i(t-1)$ has a direct influence on the same outcome measured in the present,
$Y_i(t)$.\footnote{The results of this investigation hold even if this
  assumption is dropped, or if the time dependence goes beyond the first order;
  that is, $Y_i(t-k)$ continues to influence $Y_i(t)$ even after controlling
  for $Y_i(t-1)$.}  We also assume that $X_i$ directly influences $Y_i(t)$ for
all $t$, though possibly not to the same magnitude or mechanism at each time
$t$.\footnote{The result will go through so long as $Y_i(t_0)$ is influenced by
  $X_i$ for at least one $t_0$, and for the subsequent observation $t \geq
  t_0$.}  Finally, we assume that another individual's prior outcome $Y_j(t-1)$
can directly influence $Y_i(t)$ only if $A_{ij} = 1$ --- that is, there must be
an edge present for this direct influence to occur.  We are indifferent as to
whether the observable covariates $Z_i$ have a direct influence on
$Y_i(\cdot)$, or whether it is correlated with the latent covariates $X_i$.
The upshot of these assumptions is the causal graph in Figure
\ref{fig:latent-homophily}, examination of which should make it unsurprising
that contagion, the direct influence of $Y_j(t-1)$ on $Y_i(t)$, is confounded
with latent homophily:
\begin{itemize}
\item $Y_j(t-1)$ is informative about $X_j$; 
\item $X_j$ is informative about $X_i$ when $i$ and $j$ are linked ($A_{ij}=1$); and
\item $X_i$ is informative about $Y_i(t)$.  
\end{itemize}
Thus $Y_i(t)$ depends statistically on $Y_j(t-1)$, whether or not there is a
direct causal effect of contagion present.

While this argument would appear to be loosely assembled, it can be tightened
up using the familiar rules for manipulating graphical causal models
\citep{Spirtes-Glymour-Scheines, Pearl-causality-2nd}.  $X_i$ $d$-separates
$Y_i(t)$ from $A_{ij}$.  Since $X_i$ is latent and unobserved, $Y_i(t)
\leftarrow X_i \rightarrow A_{ij}$ is a confounding path from $Y_i(t)$ to
$A_{ij}$.  Likewise $Y_j(t-1) \leftarrow X_j \rightarrow A_{ij}$ is a
confounding path from $Y_j(t-1)$ to $A_{ij}$.  Thus, $Y_i(t)$ and $Y_j(t-1)$
are $d$-connected when conditioning on all the observed (boxed) variables in
Figure \ref{fig:latent-homophily}.  Hence the direct effect of $Y_j(t-1)$ on
$Y_i(t)$ is not identifiable \citep[\S 3.5, pp. 93--94]{Pearl-causality-2nd}.

This argument is not affected by adding conditioning on $Y_i(t-1)$ or $Y_j(t)$,
as that does not remove the confounding paths.  Nor does adding conditioning on
$Z_i, Z_j$ remove the confounding.  Nor is the situation helped by allowing
$A_{ij}$, or indeed $X$, to vary over time, as is readily verified by drawing
the appropriate graphs.  Finally, adding a {\em third} individual to the graph
would not help: even if they were, say, assumed to be linked to $i$ but not $j$
or vice versa, $Y_i(t) \leftarrow X_i \rightarrow A_{ij}$ and $Y_j(t-1)
\leftarrow X_j \rightarrow A_{ij}$ would remain confounding paths.

\begin{figure}
\begin{center}
  $a$\resizebox{0.48\textwidth}{!}{\includegraphics{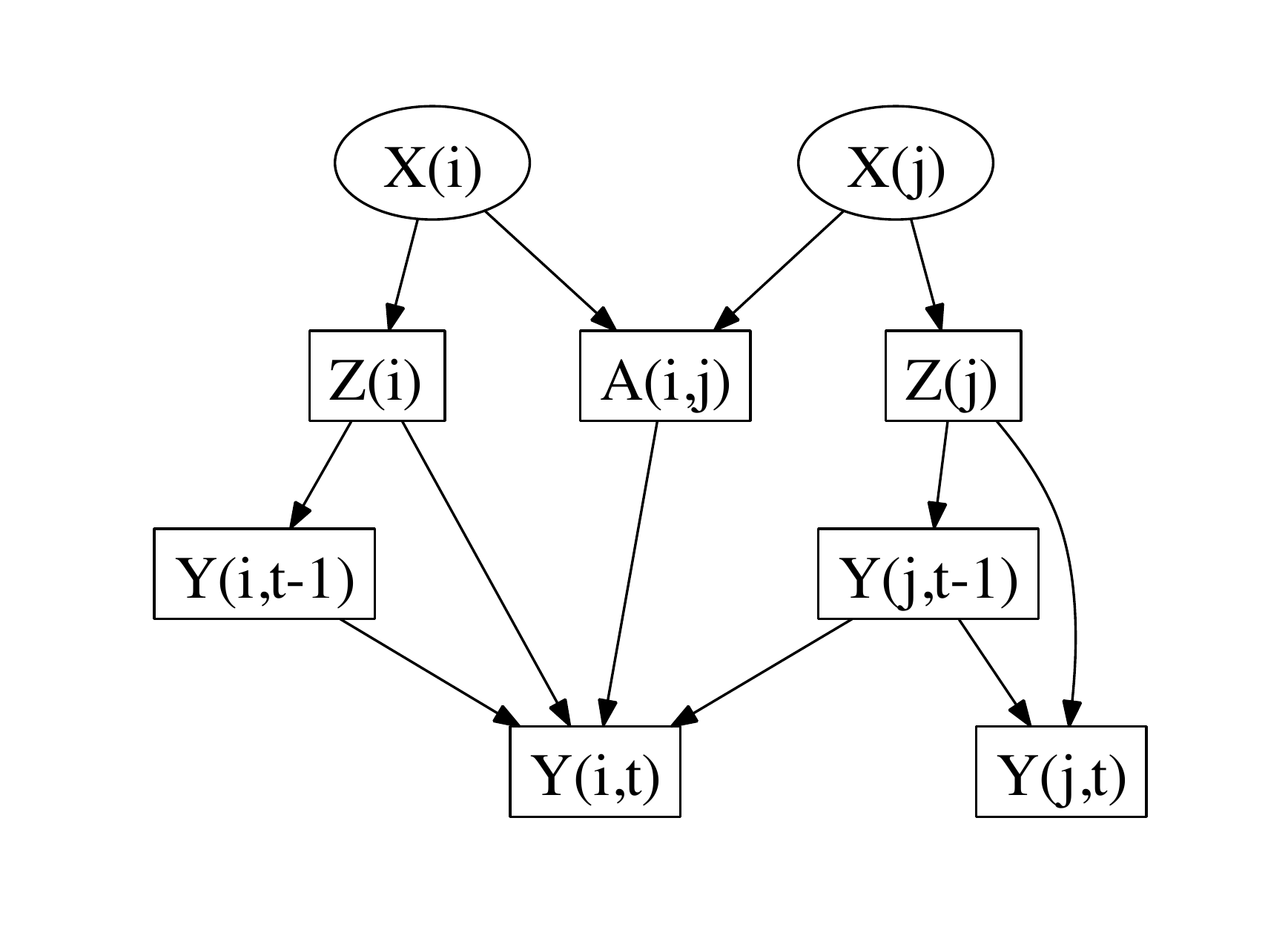}}
  $b$\resizebox{0.48\textwidth}{!}{\includegraphics{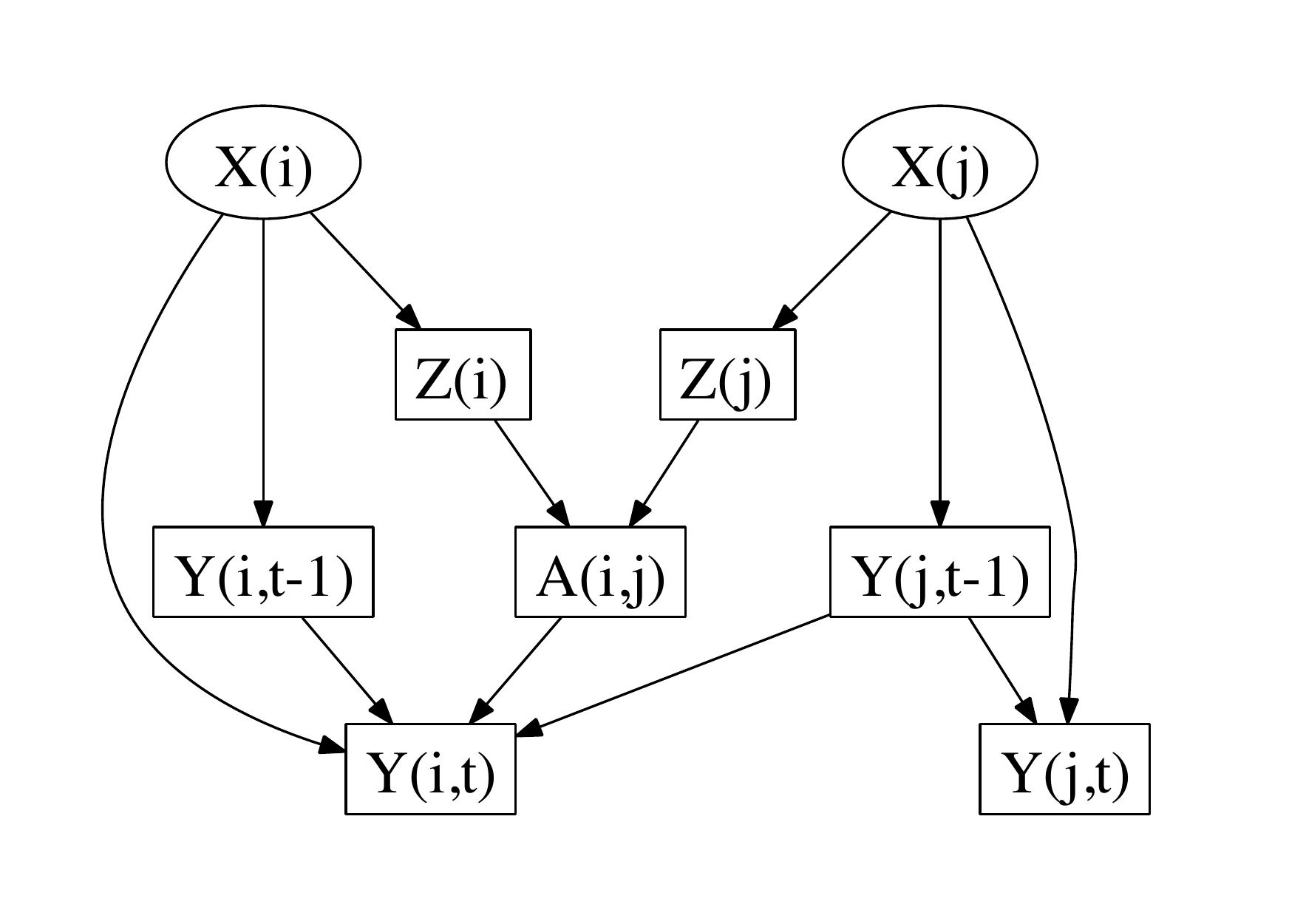}}
\end{center}
\caption{Modifications of the causal graph shown in Figure
  \ref{fig:latent-homophily}, in which observable covariates ($Z$) conveys
  enough information about $X$ that contagion effects are unconfounded with
  latent homophily.  In $a$ (left), $Z$ carries all of the causal effect from
  $X$ to the observable outcome $Y$; in $b$ (right), $Z$ carries all of the
  effect from $X$ to the social network tie $A$.}
  \label{fig:observable-control}
\end{figure}

How then might we get identifiability?  It may be that very stringent
parametric assumptions would suffice, though we have not been able to come up
with any which would be suffice\footnote{In particular, making all of the
  relations between continuous variables in Figure \ref{fig:latent-homophily}
  linear, with independent noise for each variable, is {\em not} enough --- the
  confounding path continues to prevent identifiability even in a linear
  model.}  Otherwise, we must keep $X$ from being latent, or, more precisely,
either the components of $X$ that influence $Y$ must be made observable (Figure
\ref{fig:observable-control}a), or those parts of $X$ which influence the
social tie formation $A$ (Figure \ref{fig:observable-control}b).  In either case
the confounding arcs go away, and the direct effect of $Y_j(t-1)$ on $Y_i(t)$
becomes identifiable.\footnote{\citet{Elwert-Christakis-wives-and-ex-wives} is
  another interesting approach.  In effect, they introduce a third node, call
  it $k$, where they can assume that $Y_i$ is not influenced by $Y_k$, but the
  homophily is the same.  Estimating the apparent influence of $Y_k$ on $Y_i$
  then shows the extent of confounding to due purely to homophily; if $Y_i$ is
  more dependent than this on $Y_j$, the excess is presumably due to actual
  causal influence.}  It is noteworthy that the most successful attempts at
explicit modeling that handle both homophily and influence, as found in the
work of \citet{Leenders-structure-and-influence,
  Steglich-Snijders-Pearson-selection-and-influence} involves, all at once,
strong parametric (exponential-family) assumptions, plus the assumption that
observable covariates carry {\em all} of the dependence from $X$ to $Y$ {\em
  and} $A$; the latter is also implicitly assumed by the matching methods of
\citet{aral2009distinguishing}.

Whether we face the unidentifiable situation of Figure
\ref{fig:latent-homophily}, or the identifiable case of Figure
\ref{fig:observable-control}, currently depends upon subject-matter knowledge
rather than statistical techniques. It may be possible to adapt algorithms,
such as those in \citet{Spirtes-Glymour-Scheines}, to detect the presence of
influential latent variables.  Some new methodological work would be required,
however, since all such algorithms known to us rely strongly on having a supply
of independent cases, and social networks are of interest precisely because
individuals, and even dyads, are not independent.

\subsection{The Argument from Asymmetry}
\label{sec:asymmetry}

A clever argument for the presence of direct influence was introduced by
\citet{Christakis-Fowler-spread-of-obesity}.  By focusing on unreciprocated
directed edges --- pairs $(i,j)$ where $A_{ij} = 1$ but $A_{ji} = 0$, so that
$j$'s prior outcome can be said to influence $i$'s present, but not $i$'s prior
outcome on $j$'s present --- one can consider the distributions of the outcomes
conditional on their partner's previous outcome, $Y_i(t)|Y_j(t-1)$ and
$Y_j(t)|Y_i(t-1)$ (though other observable covariates $(Z_i, Z_j)$ may also be
conditioned on.)  An asymmetry here, revealed by the difference in the
corresponding regression coefficients, might then be due to some influence
being transmitted along the asymmetric edge, and not due to external common
causes (such as a new fast food restaurant) or other behaviours attributable to
latent characteristics.

This idea has considerable plausibility and has been picked up by a number of
other authors \citep{flickr-tagging-influence,bramoulle2009ipetsn}, who have
shown that it works as a test for direct influence in some models.  However, we
show that the argument can break down if two conditions are met: first, the
influencers (the $j$ in the pair) differ systematically in their values of $X$
from the influenced (the $i$), and, second, different neighborhoods of $X$ have
different local (linear) relationships to $Y$. As previously mentioned, the
most successful claims of simultaneous accounting of these phenomena require
strong parametric assumptions, and our demonstration shows that even
assumptions of linearity may be too strong for this sort of data.

To illustrate this claim, we present a toy model of a network with latent
homophily on an $X$ variable that controls an observable time series $Y$ at
multiple points, but with no direct influence between values of $Y$ for
different nodes. We present this as a multi-step time series to approximate the
scenario of \citet{Christakis-Fowler-spread-of-obesity}, so that we can add the
two most recent time steps of the alter's expression into the
regression.\footnote{The method in \citet{Christakis-Fowler-spread-of-obesity}
  uses a ``simultaneous'' regression set-up, including $Y_j(t)$ as a predictor
  of $Y_i(t)$ as well as a previous time point $Y_j(t-1)$.  Treated at face
  value, this can produce an incoherent probability distribution for the
  evolution of the system \citep{lyons2010semvfsa}, as well as implying a
  scarcely-comprehensible notion of simultaneous causation (rather than coupled
  behavior or feedback); this can be somewhat salvaged by considering it as an
  observation that shares information from the ``$t$ minus one-half'' time
  point, as well as picking up any coupled behaviour at time $t$.}  We also
note that there is no ``coupled evolution'' of two nodes' outcomes due to an
exogenous common cause, one of the stated purposes of the asymmetry
test. Despite the lack of direct interaction, it is possible to predict $Y_i$
at time $t$ from the value of $Y$ at its neighbors for times $t-1$ and $t-2$,
and these relations are asymmetric across unreciprocated edges.

First we present the formation of the network, which contains $n$ individuals
(nodes), and each node $i$ has a scalar latent attribute $X_i \sim
\mathcal{U}(0,1)$, which are generated independently. We generate an underlying
undirected network (a potential friendship pool) where such an edge forms
between $i$ and $j$ with probability equal to
$\mathrm{logit}^{-1}(-3|X_i-X_j|)$, so that edges are more likely to form
between individuals with similar values of $X$. Each individual $i$ then
nominates their ``declared'' friendships from these neighbors, naming $j$ with
probability proportional to $\propto \mathrm{logit}^{-1}(-|X_j - 0.5|)$ ---
individuals, whatever their own value of $X$, prefer to nominate acquaintances
closer to the median value of that trait.\footnote{Whether this is an actual
  bias in the social network formation process, or merely a part of the process
  recording the network, does not matter.  Also, results would work equally
  well if ties were biased towards extreme rather than central values of $X$,
  for multivariate latent traits, and so forth.}  For this demonstration, as in
the data sets used in
\citet{Christakis-Fowler-spread-of-obesity,Fowler-Christakis-happiness}, each
individual $i$ declares one friend, though the results hold for greater numbers
of nominations. This produces the sociomatrix/adjacency matrix $A$, where
$A_{ij}=1$ signifies that individual $i$ has nominated $j$ as a ``friend''.

Second, we establish the time trends of the observable outcomes $(Y_i(t=0),
Y_i(t=1))$:
\begin{itemize}
\item At time $t=0$, we set $Y_i(0) = (X_i-0.5)^3 + \mathcal{N}(0,(0.02)^2)$, a
  nonlinear assignment of outcome attributes.
\item For time $t=1$, we set $Y_i(1) = Y_i(0)+0.4X_i+\mathcal{N}(0,(0.02)^2)$,
  so that the trend is greater for those individuals with higher values of the
  latent attribute.
\item For time $t=2$, we set $Y_i(2) = Y_i(1)+0.4X_i+\mathcal{N}(0,(0.02)^2)$,
  repeating the trend.
\end{itemize}

\begin{figure}
\begin{center}
\resizebox{0.75\textwidth}{!}{\includegraphics{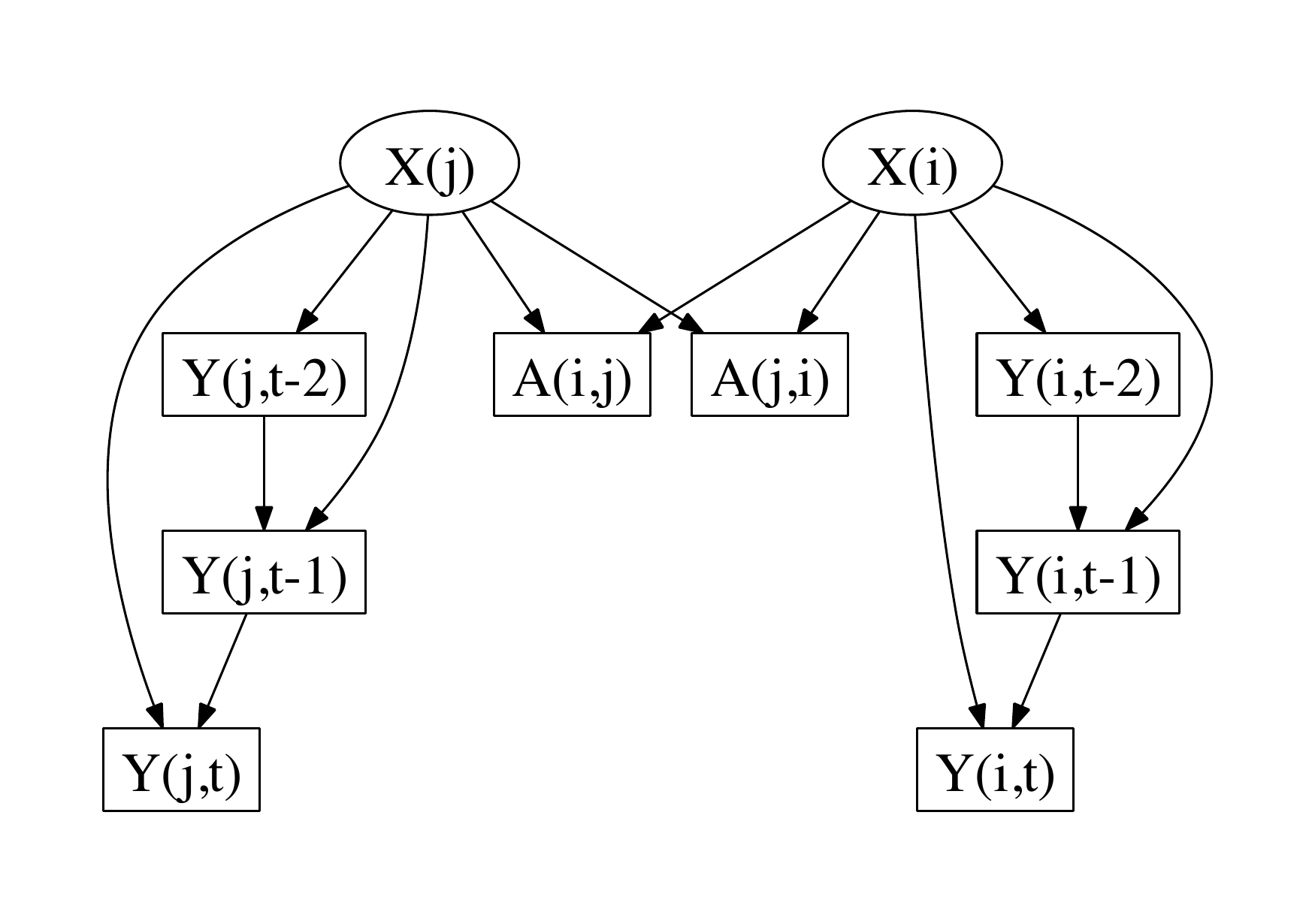}}
\end{center}
\caption{Graphical causal model for our simulation study in section
  \ref{sec:asymmetry}.  Here, unlike Figure \ref{fig:latent-homophily}, there
  are no arrows from $(Y_j(t-2), Y_j(t-1))$ to $Y_i(t)$ --- the former outcomes
  for the ``alter'' are not, in reality, a cause of the latter for the ``ego'',
  and the relationships of the $Y_j$ and $Y_i$ time series are symmetrical.  As
  we show in the text, however, not only $Y_{i}(t)$ predictable from
  $Y_j(t-1)$, but the relationship is asymmetric when social network ties are
  unreciprocated, i.e., $A_{ij}=1$ but $A_{ji}=0$.}
\label{fig:asymmetry-model}
\end{figure}

Figure \ref{fig:asymmetry-model} is the graphical model for the actual causal
structure of our simulation at three time points.

We simulate a network of fixed size ($n=400$) from this model and estimated the
linear model
\begin{eqnarray*}
Y_i(2) & = & \alpha + \beta_1 Y_i(1) + \beta_2 \sum_j A_{ij}Y_j(1) + \beta_3 \sum_j A_{ji}Y_j(1) + \\
& & \beta_4 \sum_j A_{ij}Y_j(0) + \beta_5 \sum_j A_{ji}Y_j(0)+ \epsilon_i ~,
\end{eqnarray*}
so that $\alpha$ represents the intercept term and $\beta_1$ represents the
autocorrelation; $\beta_2$ is the effect of the nominee's status at time $t=1$
on the nominator, and $\beta_3$ is the converse, the network effect if $i$ was
nominated by $j$, at time $t=1$; $\beta_4$ and $\beta_5$ are those same
coefficients for the outcome at time $t=0$. This was replicated 5000 times,
with the latent variables, time series and network regenerated in each
replication.

\begin{figure}
\begin{center}
$\begin{array}{cc} \includegraphics[width=0.45\textwidth]{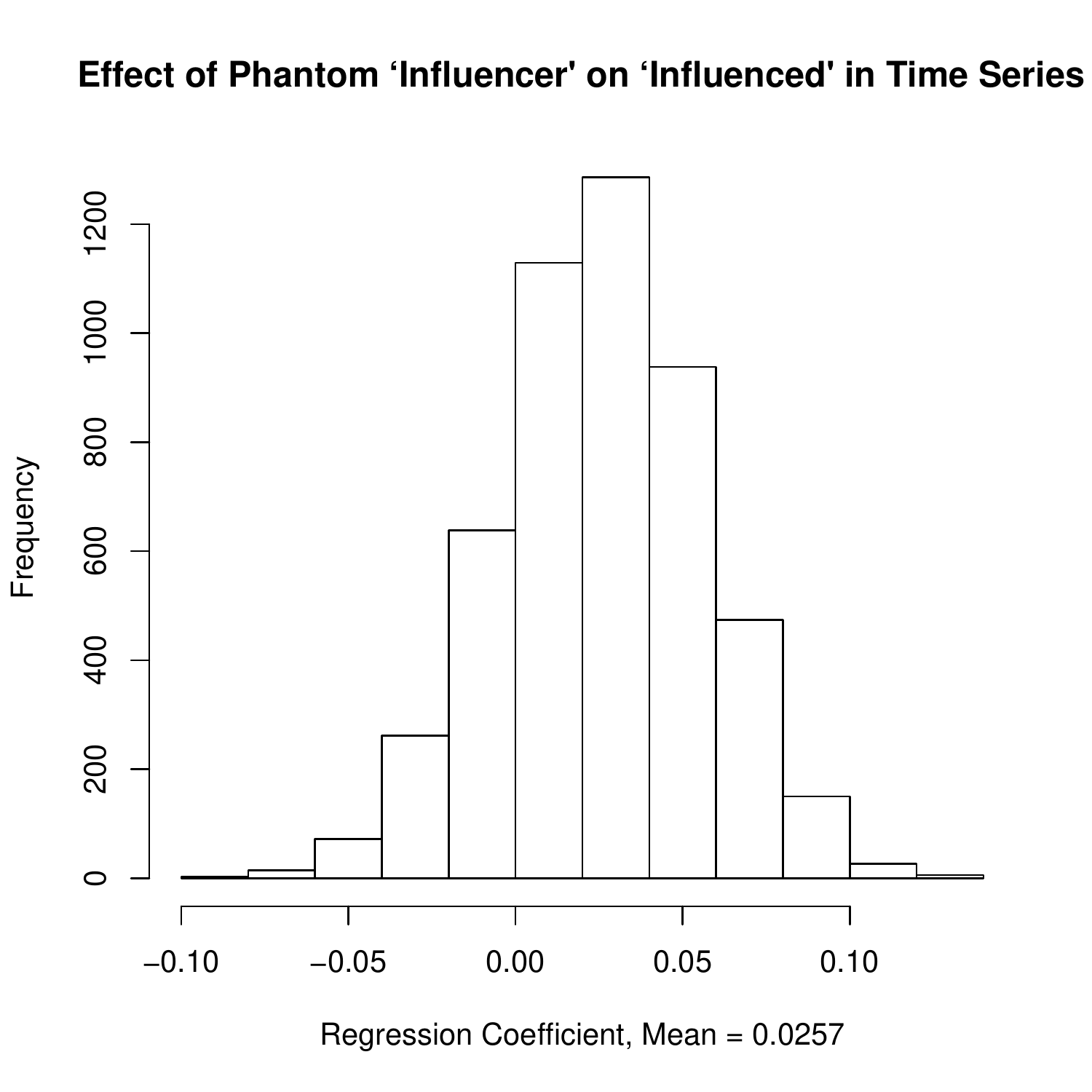} & 
\includegraphics[width=0.45\textwidth]{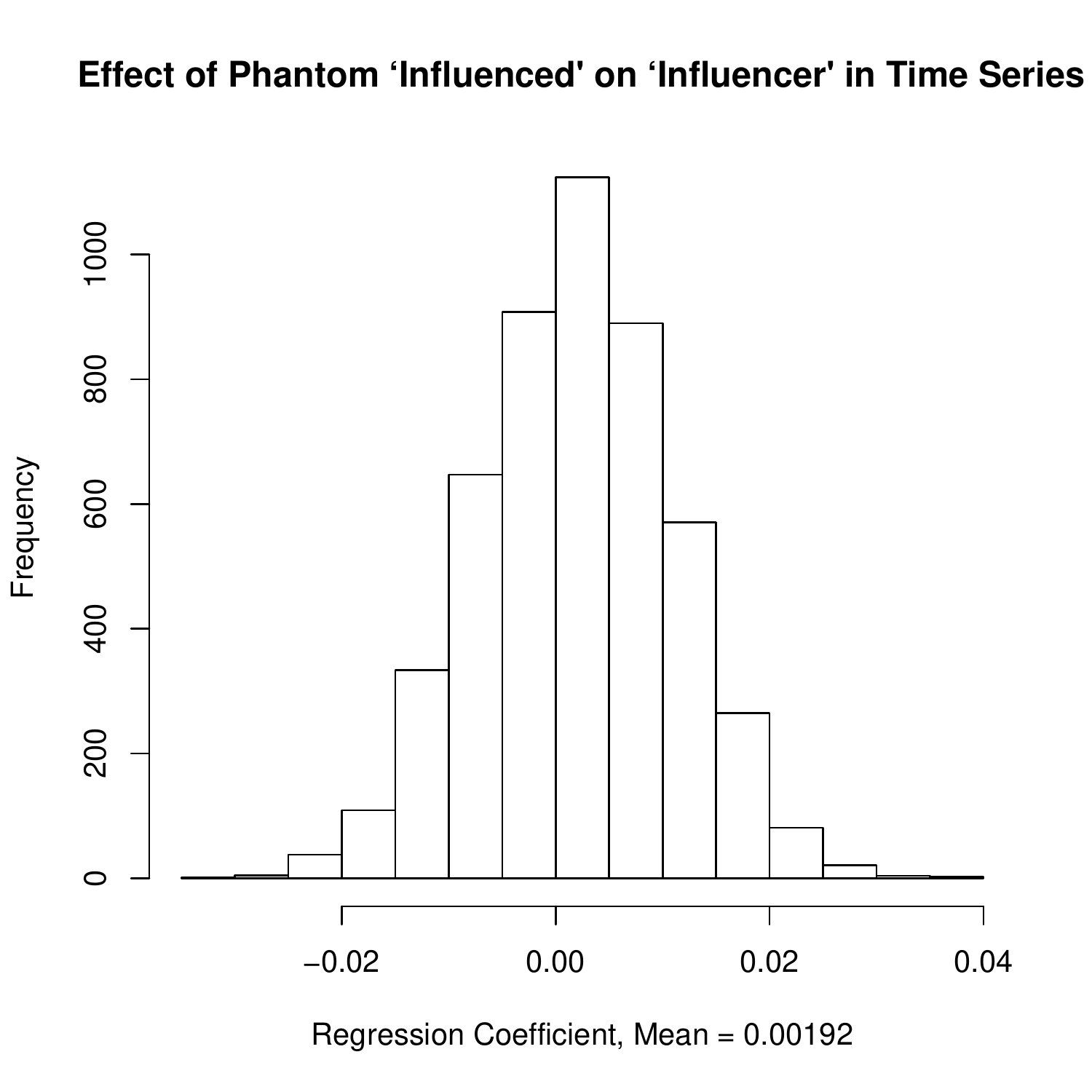} \\
\includegraphics[width=0.45\textwidth]{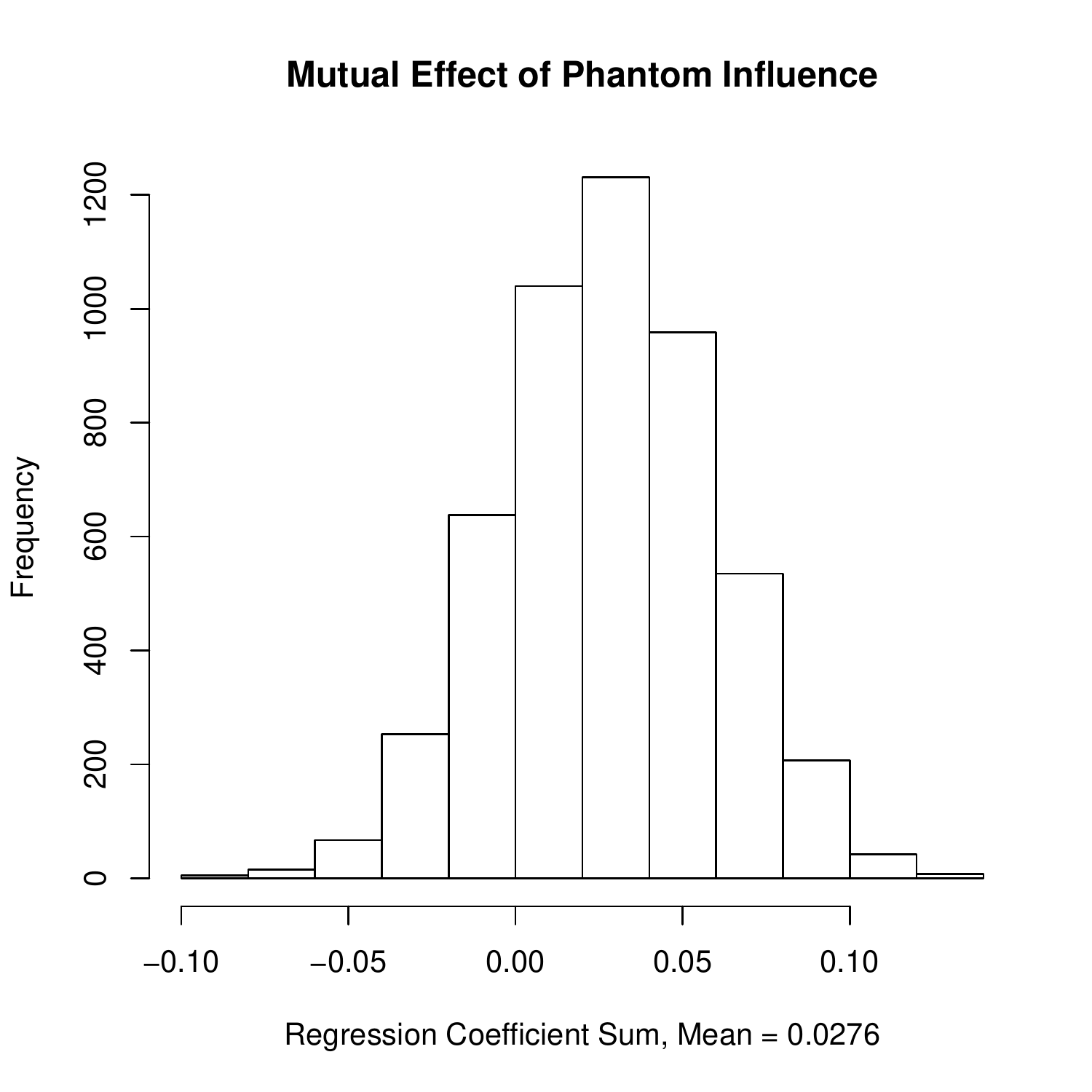} & 
\includegraphics[width=0.45\textwidth]{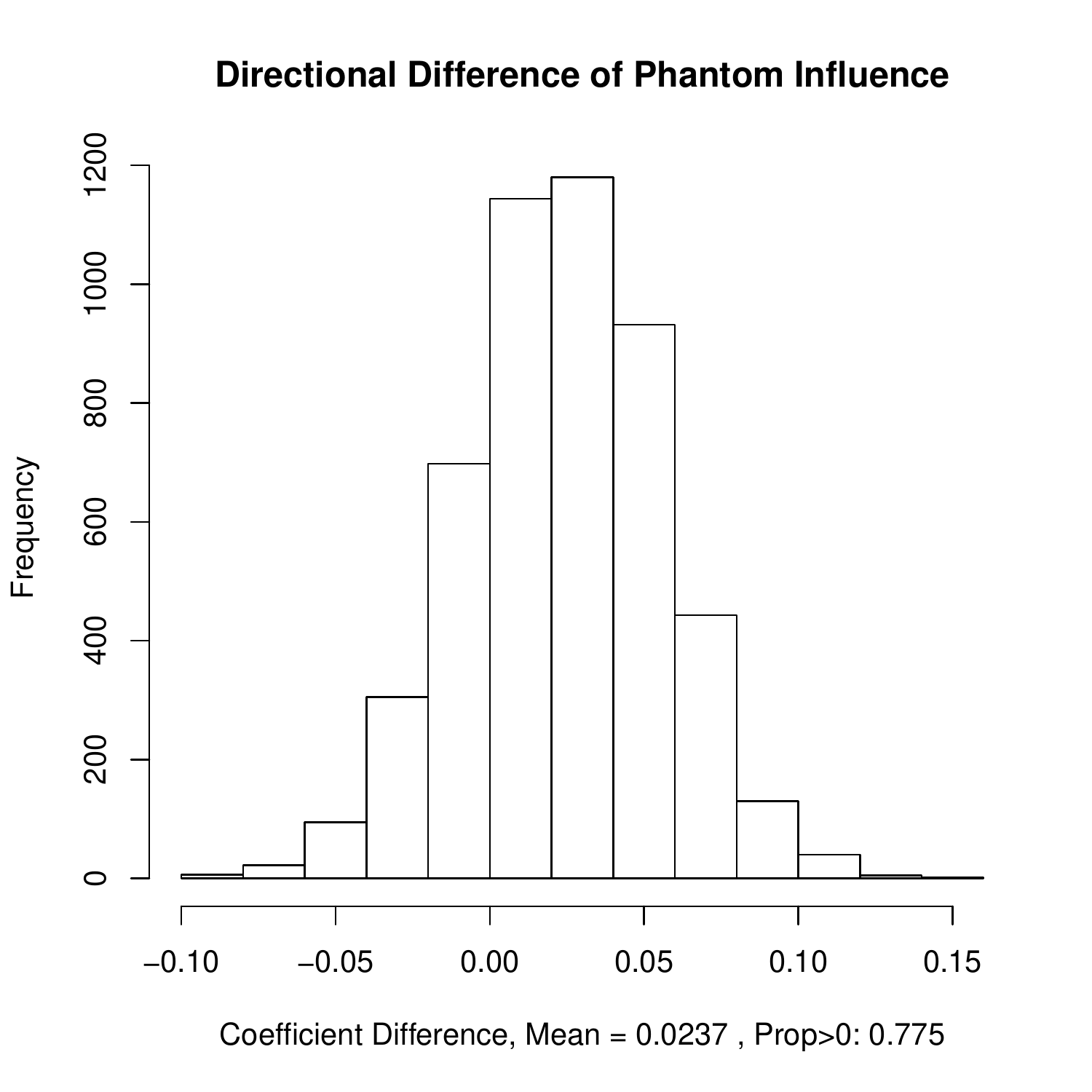} \end{array}$
\end{center}
\caption{Results for a toy model where a latent variable causes spurious
  time-dependent network effects. Clockwise from the top left: a) The estimate
  for $\beta_2$, the effect in the expected direction of influence. b) The
  estimate for $\beta_3$, the effect in the opposite direction of influence
  (from the namer to the named). c) The sum of the estimated effects,
  indicating that the effect for a mutual tie (in which each respondent names
  the other) is greater than either the expected or opposite unreciprocated tie
  effect. d) The normalized difference between directional effects is clearly
  greater than zero on balance (in roughly 77\% of simulations), suggesting
  that the asymmetry in coefficient estimates can be produced without contagion
  and falsely detected by $t$-tests on the
  difference. \label{fig:timeseriesmodel}}
\end{figure}

Figure \ref{fig:timeseriesmodel} shows the results of these simulations.
Figure \ref{fig:timeseriesmodel}$a$ shows the magnitude of $\beta_2$, the
coefficient of network influence; in 4010 of these 5000 trials is the estimate
less than zero despite the lack of a direct connection, in line with the
empirical results of \citet{Cohen-Cole-Fletcher}. This is also the case for
Figure \ref{fig:timeseriesmodel}$b$, showing the apparent coefficient of a
``reverse'' network effect $\beta_3$, which is smaller in magnitude. Figure
\ref{fig:timeseriesmodel}$c$ shows the sum of the two effects; this
demonstrates that the effect of a mutual tie, where $A_{ij}A_{ji}=1$, is
determined by the sum of the one-way effects and is greater than the effect of
a ``named'' tie, $A_{ij}=1$, which is greater than the effect of a ``naming''
tie, $A_{ji}=1$. This is the result of the type that was cited in
\citet{Christakis-Fowler-spread-of-obesity, Fowler-Christakis-happiness} but
produced without any network interaction.\footnote{There is also the notion of
  a ``bonus'' effect for mutual ties, $\beta_4 \sum_j A_{ij}A_{ji}Y_j(0)$,
  which could provide an additional bump for mutuality that would indicate a
  stronger tie than simply indicated by a binary specification. We leave this
  for another investigation, noting that the mutual $>$ named $>$ namer
  relation is satisfied without adding this term.}

Figure \ref{fig:timeseriesmodel}$d$ shows the difference between the ``sender''
and ``receiver'' coefficients, which would be approximately Gaussian (for a
$t$-distribution with 400 degrees of freedom) and centered at zero, if this
were the case, a $t$-test could be used to claim statistical significance in
the difference between the two effects.  It is evident from the histogram that
this null distribution is not centered at zero, and about 77\% of the sample
values are positive, even though there is really no effect.  Thus, latent
homophilous variables can produce a substantial apparent contagion effect,
including the asymmetry expected of actual contagion.

The parameter values in this model were not chosen to maximize either the
apparent contagion effect or its asymmetry, merely to demonstrate their
presence. As well, we note that controlling for additional past values of the
property for each node reduces the imbalance in magnitude, while it still
remains statistically significant; as we show in Section \ref{sec:bounds}, this
is not the end of the story if we cannot find a bound for this asymmetry.

Additionally, it may seem unlikely that these conditions may exist on
unobserved variables in the system, but this still places the burden on the
investigator to pursue as many possible latent factors as may be present --- an
extremely onerous task in a multi-decade observational study --- or to work
exclusively with experimental data, such as in the recent work of
\citet{Fowler-Christakis-2010cbchsn}.

\section{How Contagion and Homophily Look Like Causation At The Individual Level}
\label{sec:contagion-fakes-causation}

We would be remiss if we gave the impression that it is only investigators who
actually take network structure into account who have problems.  In this
section, we show that a very common kind of use of survey data, namely that
relating individual's choices (cultural, political, economic, etc.) to their
long-term stable traits, is {\em also} confounded in the presence of homophily
and contagion.  Continuing the spirit of Section \ref{sec:asymmetry}, we
present another toy model in which regressions of choices on traits produce
significant non-zero coefficients that are {\em solely} due to this
confounding.\footnote{Preliminary versions of these results appeared in
  \citet{CRS-social-media}, and as long ago as 2005 at
  \url{http://bactra.org/notebooks/neutral-cultural-networks.html}.  We
  understand from a presentation by Prof.\ Miller McPherson that he and
  colleagues have been working on parallel lines, and will soon publish a
  demonstration that biases of this sort can be quite substantial even for the
  canonical General Social Survey (M. McPherson, ``Social Effects in Blau
  Space'', presentation at MERSIH 2, 14 November 2009).}

It should be emphasized that there is a long tradition within social science of
distinguishing long-term, hard-to-change aspects of social organization and
individuals' place in it, from more short-term, malleable aspects which show up
in behavior and choices.  As Ernest \citet{Gellner-cause-and-meaning} put it,
``Social structure is who you can marry, culture is what you wear at the
wedding.''  The long-standing theoretical presumption, common to all the
classical sociologists (even, in his own way, to Max Weber), and going back
through them to Montesquieu if not beyond \citep{Aron-main-currents}, is that
social structure {\em explains} culture, or that the latter reflects the
former; in many versions, culture is an {\em adaptation} to social structure.
This intuition is alive and well through the social sciences, the humanities,
and among lay people.  Many of these accounts have considerable plausibility,
though since they conflict with each other they cannot all be true.  However,
aside from casual empiricism, the {\em evidence} for them consists largely of
correlations between cultural choices and social positions, demonstrations that
the superstructure can be predicted from the base.  Famously, for instance,
\citet{Bourdieu-distinction} attempts to do this for survey data.

We do not wish to assert that social position is {\em never} a cause of
cultural choices; like everyone else, we think that it often is.  The issue,
rather, is the evidence for such theories, and in particular for the magnitude
of such effects.

\begin{figure}
\begin{center}
  \resizebox{0.5\textwidth}{!}{\includegraphics{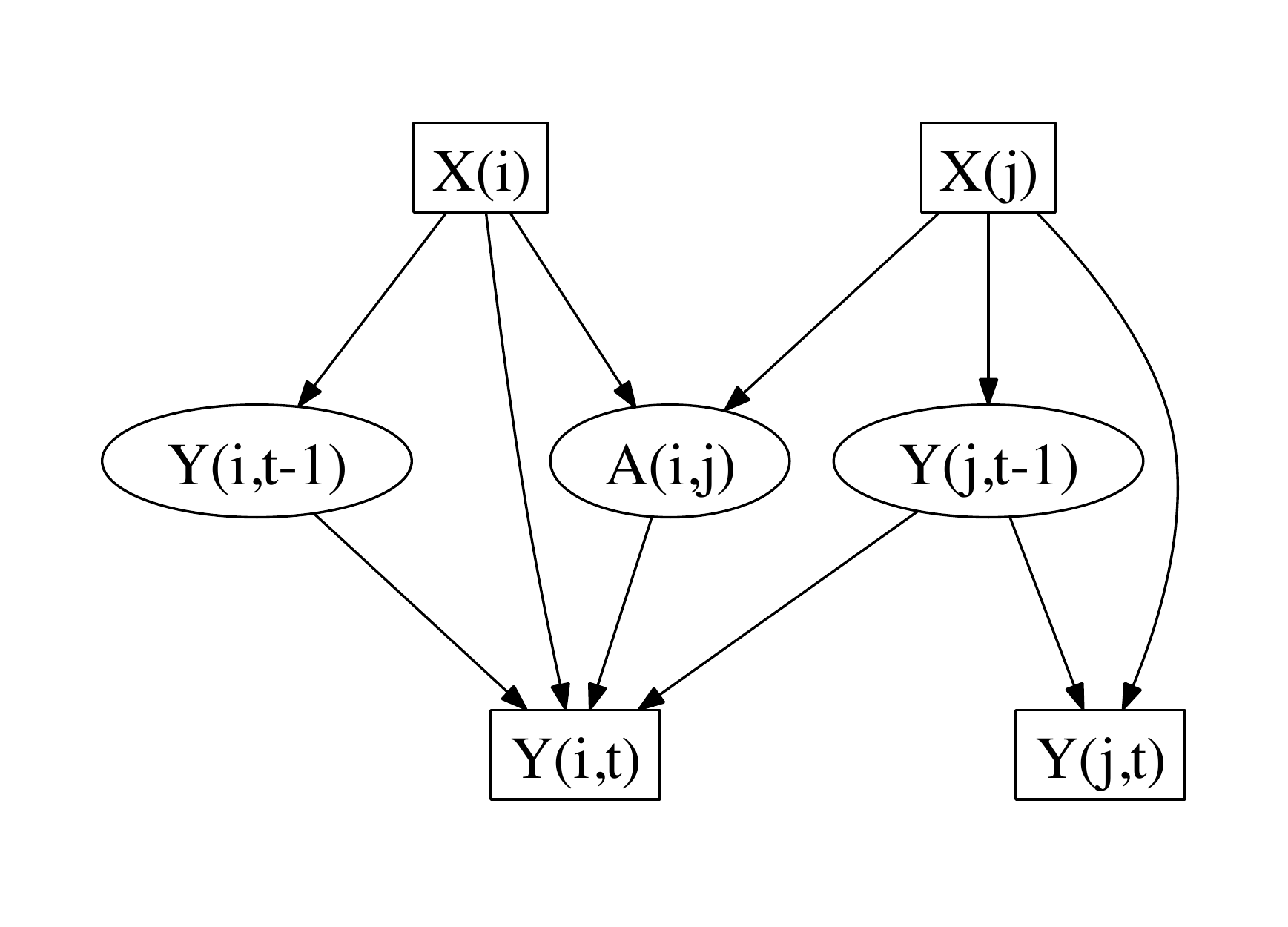}}
\end{center}
  \caption{Typical situation in surveys linking cultural choices to social
    traits when homophily and influence exist.}
  \label{fig:survey-regression2}
\end{figure}

\subsection{Simulation Model}
\label{sec:voter-model}

We work with what is, frankly, a toy model of contagion (though, see footnote
\ref{fn:social-influence-models} below).  There are $n$ individuals connected
in an undirected social network.  Each individual $i$ has an observed trait
$X_i$ which is an unchanging variable; in our examples, this will be binary.
The network is homophilous on this trait, so that individuals with the same
value of $X$ are more likely to be connected.  Individuals also have a
time-varying choice variable $Y_i(t)$, which again we will take to be
binary. The initial choices, $Y_i(0)$, are set by flipping a fair coin (i.e. an
unbiased Bernoulli process), and are therefore independent of the traits $X_i$.

\begin{figure}
\begin{center}
\resizebox{0.5\textwidth}{!}{\includegraphics{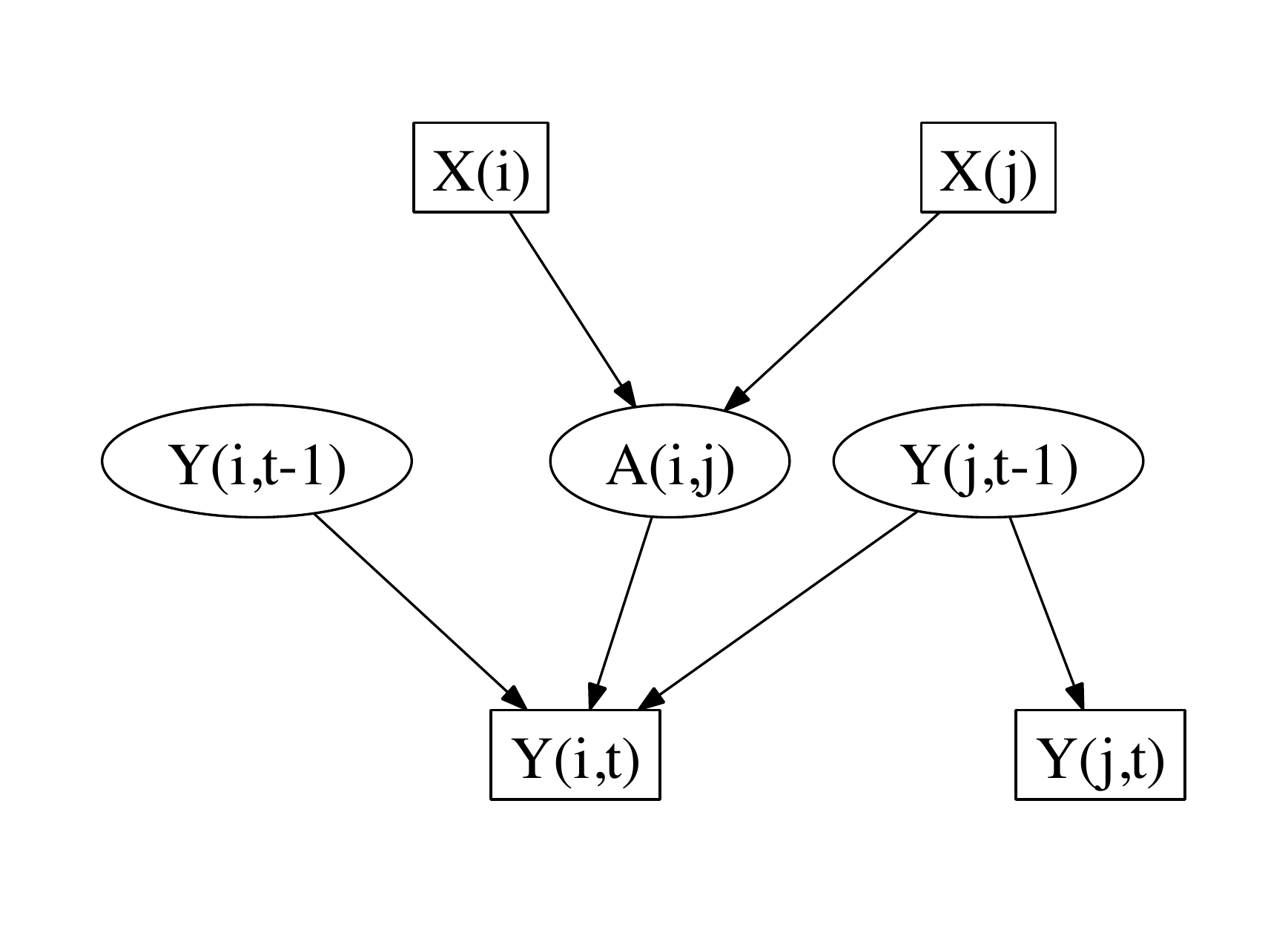}}
\end{center}
\caption{Graphical model showing the causal structure of the model simulated in
  Section \ref{sec:voter-model}; cf.\ Figure
  \ref{fig:survey-regression2}. Notice that here, the persistent traits $X$
  have no direct causal influence on the choices $Y$.  As we show, however,
  diffusion of choices along homophilous ties creates states where $Y$ can be
  predicted from $X$.}
\label{fig:voter-model-graph}
\end{figure}

Choices evolve as follows: at each time $t$, we pick an individual $I_t$,
uniformly at random from $i \in \{1, ..., n\}$, independently of all prior
events.  This individual then picks a neighbor, again uniformly at random, $J_t
\in \{j:A_{I_{t}j}=1\}$, and either, with very high probability, copies their
choice, so that $Y_{I_t}(t) = Y_{J_t}(t-1)$, or, with very low probability,
assumes the opposite choice, for $Y_{I_t}(t) = 1-Y_{J_t}(t-1)$; all other
individuals retain their previous choices. This process repeats for each time
step.  Figure \ref{fig:voter-model-graph} shows the causal structure.

This random copying model is, of course, a drastic oversimplification of actual
processes of transmission and influence, which have been extensively studied in
social psychology and allied fields since the 1920s
\citep{Bartlett-remembering, Sperber-explaining,
  Huckfeldt-Johnson-Sprague-political-disagreement,
  Friedkin-social-influence}.\footnote{Notice that the expected value of
  $Y_{I_t}(t+1)$ is just the mean of $Y_j(t)$ for the $j$ neighboring $I_t$.
  The expected value of $Y_{i}(t+1)$ for all $i$ is thus a weighted average of
  $Y_i(t)$ and the mean of their neighbors.  At the level of expectations,
  then, this process belongs to the family of linear social influence models
  used in, e.g.,
  \citet{Friedkin-social-influence}. \label{fn:social-influence-models}.}
However, not only is it adequate to demonstrate the existence of the phenomenon
we are concerned with, its very abstraction helps indicate just how robust the
problem is.

Probabilistically, the vector $Y(t)$ is a Markov chain, specifically a variant
of the ``voter model'' of statistical mechanics on a graph
\citep{Liggett-particle-systems, Sood-Redner-voter-model}; the minor addition
of low-frequency noise (doing the opposite of the selected neighbor) keeps
the homogeneous configurations (where $Y_i$ is constant over $i$) from being
absorbing states, but has little influence on the medium-run behavior we are
concerned with.

\begin{figure}
  \begin{center}
    \resizebox{!}{0.4\textheight}{\includegraphics{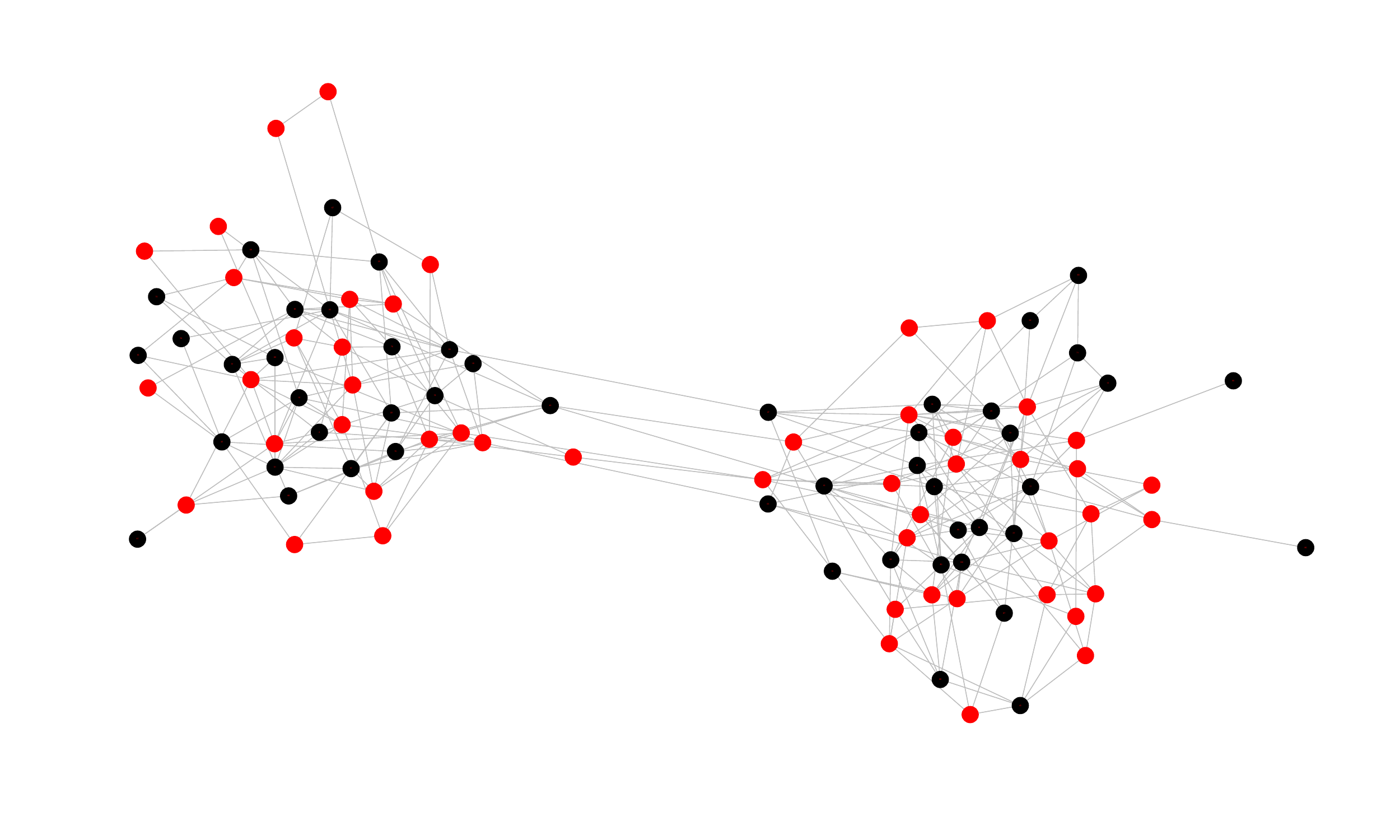}}
    \resizebox{!}{0.4\textheight}{\includegraphics{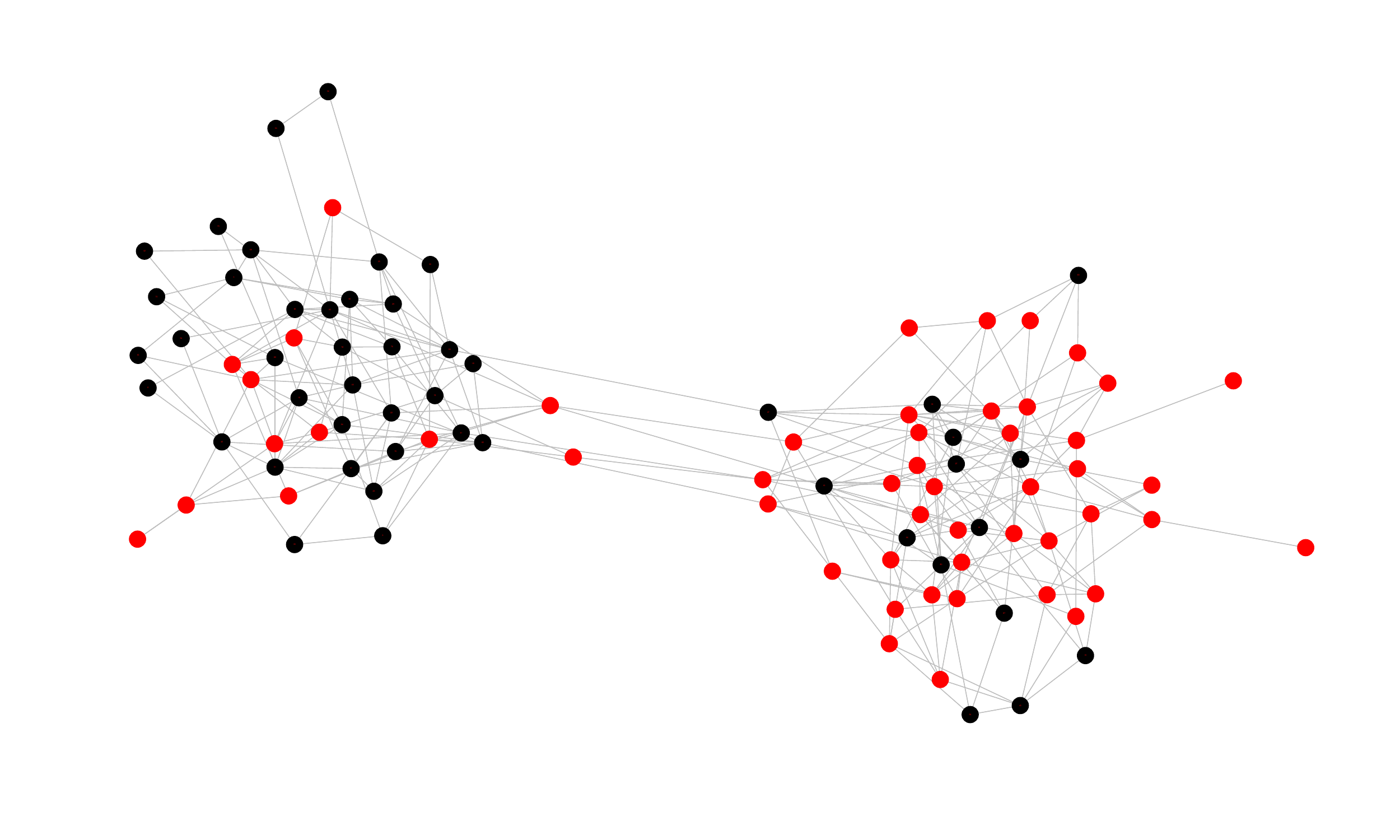}}
  \end{center}
  \caption{An illustration of the diffusion process on a network with
    homophilous ties; members of the left and right clusters have attribute
    values of 0 and 1 respectively. Initially (top), there is very little
    detectable similarity between choices within each cluster; however, after a
    few hundred time steps (bottom), there is a clear association between trait
    and cluster caused entirely by the diffusion along homophilous ties.}
\label{fig:network-evolutions}
\end{figure}

\begin{figure}
  \begin{center}
   \resizebox{0.49\textwidth}{!}{\includegraphics{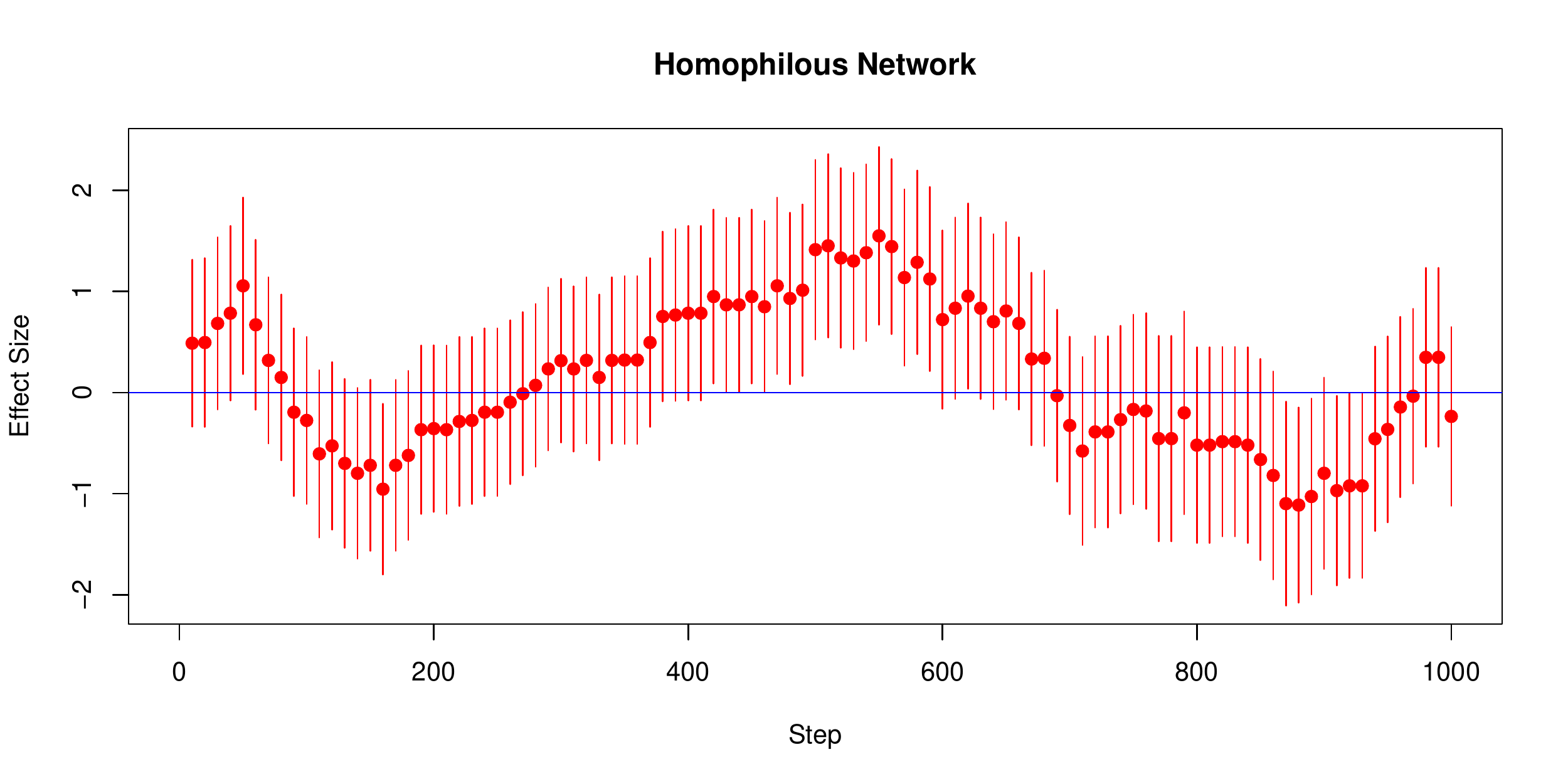}}
   \resizebox{0.49\textwidth}{!}{\includegraphics{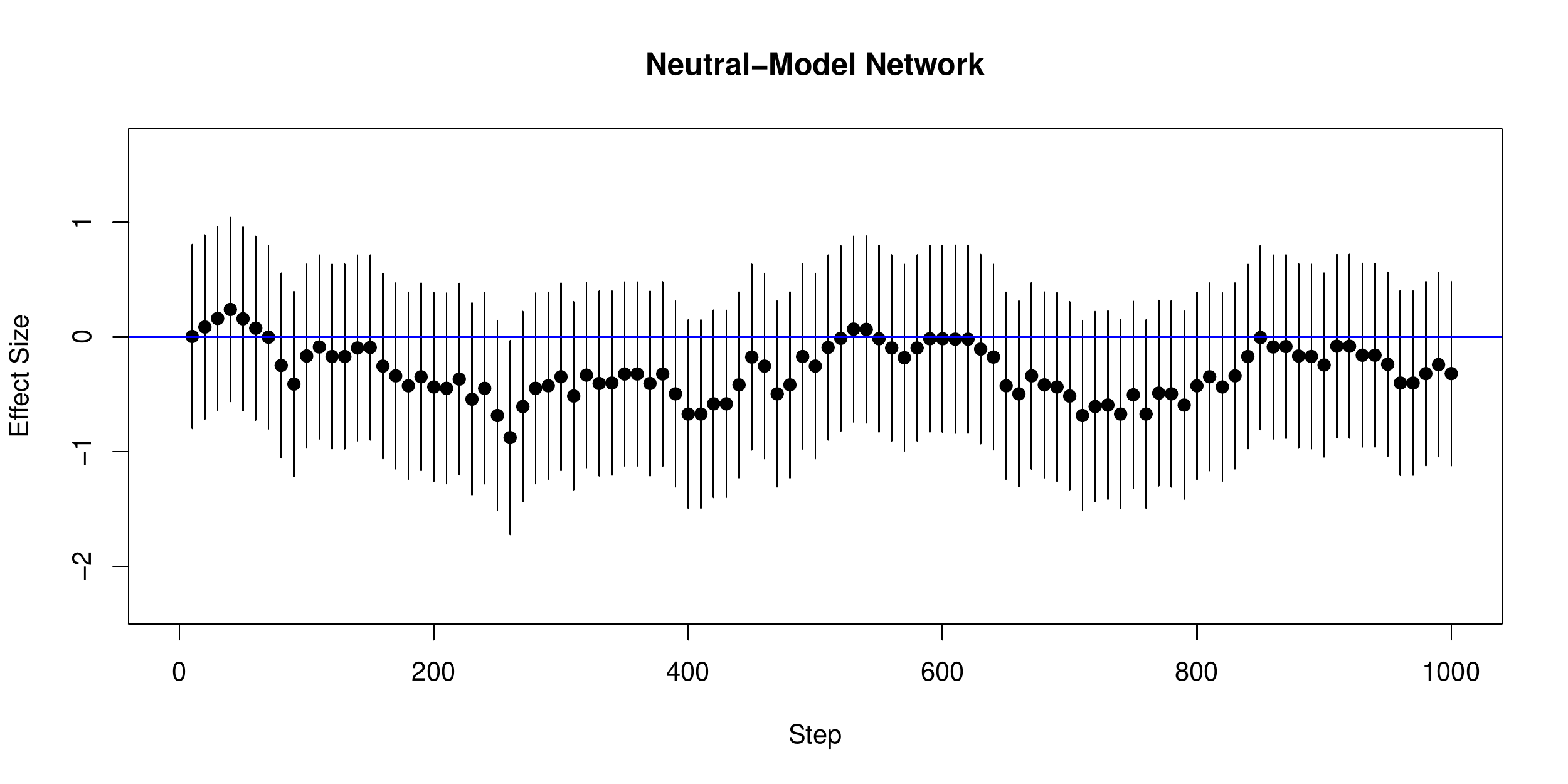}}
  \end{center}
  \caption{Coefficient estimates for logistic regressions of choice on trait as
    functions of time. Error bars represent 95\% confidence intervals on each
    run, independent of all others.  {\em Left}: the evolution in a homophilous
    network; in this run of the simulation, the coefficient first becomes
    negative and statistically significant, then becomes {\em positive} and
    significant, purely due to diffusion along homophilous ties, before
    returning to a state of negative significance.  {\em Right}: a
    corresponding series of estimates in a network where ties form
    independently of traits; deviations from neutrality are much smaller.}
\label{fig:network-correlations}
\end{figure}

Figure \ref{fig:network-evolutions} shows a typical evolution of this model. In
the top image at the initial state of the system, there are two clusters based
on social traits $X$, but the individual cultural choices (colors represent
values of $Y$) are independent of these traits. The bottom image shows the same
network and configuration after 3000 updates.  Now, even by eye, it is clear
that one of the choices has become associated with one of the social types.

This can be confirmed more quantitatively by doing a logistic regression of
choice on trait (Figure \ref{fig:network-correlations}) at several points
during the diffusion process. In this particular example, there are significant
deviations in each direction. First, the association between trait 1 and color
1 is positive and significant, and remains so for several dozen iterations;
then the diffusion reverses the association, which then becomes negative and
significant. For comparison, a network with the same average degree but no
homophilous tie formation is shown to undergo the same diffusion process but
with no corresponding association between choice and trait. \footnote{Note that
  the standard errors are from the isolated logistic regression at each time
  point; when taken collectively, the errors in the effect size would be
  different. Our point remains that this would be the effect size estimated if
  the time evolution were not properly accounted for.}

Intuitively, the copying process tends to make neighbors more similar to each
other; Ian's choice can be predicted from Joey's choice.  On regular lattices,
this mechanism causes the voter model to self-organize into
spatially-homogeneous domains, with slowly shifting boundaries between them
\citep{Cox-Griffeath-voter-model}.  A similar process is at work here, only,
owing to the assortative nature of the graph, neighbors tend to be of the same
social type.  Hence social type is an indirect cue to network neighborhood, and
accordingly predicts choices.

To summarize, this ``neutral'' process of diffusion, together with homophily,
is sufficient to create what looks like a causal connection between an
individual's social traits and cultural choice.  This is because individuals'
choices are {\em not} independent conditional on their traits, as is generally
assumed in, e.g., survey research; diffusion creates the observed
dependence.\footnote{It should be clarified here that the problem is {\em not}
  the ecological fallacy, or a red-state/blue-state issue,
  \citep{Gelman-et-al-red-state-blue-state} since the simulation is not
  aggregating any data.}

This demonstration shows that it is difficult to argue that, for example, being
of type 0 is an {\em indirect} cause of picking the color black as opposed to
red, since even within a single run of the model the association can be seen to
reverse. Put another way, differences in social types are at most related to
differences in choices, not to the actual content of those choices.

\section{Constructive Responses}
\label{sec:positives}

To sum up the argument so far, we have shown that latent homophily together
with causal effects from the homophilous trait cannot be readily distinguished,
observationally, from contagion or influence, and that this remains true even
if there is asymmetry between ``senders'' and ``receivers'' in the network.  We
have also shown that the combination of homophily and contagion can imitate a
causal effect of the homophilous trait.  It requires little extra to see that
contagion, plus a causal influence of the contagious trait, yields a network
that contains the appearance of homophily.  Thus, given any two of homophily,
contagion, and individual-level causation, the third member of the triad seems
to follow.

We realize that these results appear to wreck the hopes on which many
observational studies of social networks have rested.  It would be nice to
think that something could, nonetheless be salvaged from the ruins.  The
``easy'' solution is to use expert knowledge of the system to identify all
causally relevant variables, measure a sufficient set of them, and adjust for
them appropriately
\citep{Morgan-Winship-counterfactuals,Pearl-causality,Spirtes-Glymour-Scheines}.
Since this is clearly a Utopian proposal, we sketch three constructive
responses which may be possible when dealing with network data where the causal
structure is imperfectly understood or incompletely measured.  These are to
randomize over the network, to place bounds on unidentifiable effects, and to
use the division of the network into communities as a proxy for latent
homophily.

\subsection{Identifying Contagion from Non-Neighbors}
\label{sec:contagion-from-non-neighbors}

The essential obstacle to identifying contagion in the setting of Figure
\ref{fig:latent-homophily} is that the presence or absence of a social tie
$A_{ij}$ between individuals $i$ and $j$ provides information on the latent
variable $X_i$, whether we implicity include the tie by predicting $Y_i(t)$
from the past values of neighbors $Y_{j}(t-1)$ or we explicitly add $A_{ij}$ to
the prediction model.  In the language of graphical models, conditioning or
selecting on $A_{ij}$ ``activates the collider'' at that variable. This
suggests that we would do better, in some circumstances, to construct a useful
inference by deliberately {\em not} conditioning on the social network, thereby
keeping the collider quiescent.\footnote{Thanks to Peter Spirtes and Richard
  Scheines for making this paradoxical suggestion.}  We outline this method to
demonstrate the possibility, rather than to advocate a new prescription for
solving the problem.

We can conduct the following procedure over many repeated trials:
\begin{enumerate}
\item Divide the nodes into two groups, by assigning each node to one of two
  bins with equal probability; let these groups be labelled as $J_1$ and $J_2$.
\item Let $Y_{J_1}(t)$ be the vector-valued time series obtained by collecting
  each of the $Y_{i}(t)$ for $i \in J_1$ into one object, and similarly for $Y_{J_2}(t)$.
\item Use some available mechanism to predict the time series for the first
  bin, $Y_{J_1}(t)$. from its lagged counterpart, $Y_{J_2}(t-1)$, while
  controlling for the previous time point within the first half,
  $Y_{J_1}(t-1)$.
\end{enumerate} 

By repeating this procedure, then averaging over all iterations (producing new
partitions each time), there will be a non-zero predictive ability if and only
if there is actual contagion or influence. We can see why one must average over
multiple divisions as follows.  Clearly, influence is possible between the two
halves only if there are social ties linking them.  However, there will
generally exist {\em some} way of picking $J_1$ and $J_2$ so that there are no
linking ties, and in the presence of homophily, those will tend to be divisions
of the network into parts which are unusually dissimilar in their homophilous
traits.  If we restricted ourself to values of $J_1$ and $J_2$ which did have
linking ties, we would once again be selecting on the homophilous trait and
activating colliders.

This may not be a practical method, as the statistical power of this test may
be very low --- the data have very high dimension, and the method deliberately
selects random predictors --- but it will be non-zero.

Even the random-halves test will fail, however, if we add a {\em direct} causal
effect of $X_j$ on $Y_{i}(t)$ (or one modulated by $A_{ij}$).  We omitted such
a link in Figure \ref{fig:latent-homophily} and subsequently, on the assumption
that {\em causal} effects between individuals must pass through observed
behavior $Y$, but this is a non-trivial substantive hypothesis requiring
rigorous justification.

\subsection{Bounds}
\label{sec:bounds}

In Sections \ref{sec:homophily-fakes-contagion} and
\ref{sec:contagion-fakes-causation}, we saw that certain causal effects were
not identifiable; that different causal processes could produce identical
patterns of observed associations.  As
\citet{Manski-identification-for-prediction} emphasizes, even when parameters
(such as the causal effect of $Y_j(t-1)$ on $Y_i(t)$) are observationally
unidentifiable, the distribution of observations may suffice to {\em bound} the
parameters.  (With sampled data, the empirical distribution of observations
generally provides estimators of those bounds.)  Sometimes these bounds can be
quite useful, even in the general non-parametric case.

We thus propose as a topic for future research placing bounds on the causal
effect of $Y_{j}(t-1)$ on $Y_i(t)$ in terms of observable associations,
assuming the structure of Figure \ref{fig:latent-homophily}.  If the bound on
this effect excluded zero, that would show the observed association could not
be due {\em solely} to homophily, but that some contagion must also be present.

If we keep the causal structure of Figure \ref{fig:timeseriesmodel}, assuming
that the $Y$ and $X$ variables are all jointly Gaussian and all relations
between continuous variables are linear\footnote{Note that our simulation had a
  non-linear relationship between $X_i$ and $Y_i$.} would let us employ the
usual rules for linear path diagrams \citep{Spirtes-Glymour-Scheines}.  The
standardized linear-model coefficient for regressing senders on receivers,
i.e., $Y_i(t)$ on $Y_j(t-1)$, controlling for all other observables, turns out
to be
\[
\rho[X_j, Y_j(t-1)] \rho[X_i, X_j|A_{ij}=1]\rho[X_i, Y_i(t)]
\]
where $\rho[K,L]$ is the path coefficient between $K$ and $L$ (and
$\rho[K,L|M]$ is the path coefficient given the required condition
$M$, rather than an observable that would be controlled for). Clearly, any
standardized regression coefficient can be obtained here by adjusting
path coefficients for unobserved variables $X$.  Thus a bound on the true causal
effect cannot be based on the linear regression coefficient {\em alone}, but we
hope it may still be possible to find a bound which uses more information about
the pattern of associations.

It would also be valuable --- and perhaps more tractable --- to place limits on
the magnitude of the association which could be generated solely by homophily.
Parallel remarks apply to bounding the causal effect of $X_i$ on $Y_i(t)$
assuming the structure of Figure \ref{fig:survey-regression2}; we suspect,
though merely on intuition, that this will be harder than bounding contagion
effects.

Along these lines, it would be particularly interesting to bound the degree of
asymmetry in regressions which can be generated in the absence of direct causal
influence (as in Section \ref{sec:asymmetry}).  Even though asymmetry as such
can be produced in the absence of influence or contagion, it could be that by
some standard, {\em really big} asymmetries can only plausibly be explained by
influence, so that detecting such asymmetries would be evidence for
influence. More exactly, if one can establish that in the absence of direct
influence the degree of asymmetry can be at most $\alpha_0$, and one finds an
actual asymmetry of $\widehat{\alpha} > \alpha_0$, then the hypothesis of
influence has passed a more or less severe test \citep{Mayo-error}, the
severity depending on the ease with which sampling fluctuations and the like
can push the estimated asymmetry $\widehat{\alpha}$ over the threshold when the
``true'' asymmetry (in the population or ensemble) was below it.

\subsection{Network Clustering}
\label{sec:clustering}

Since the problems we have identified stem from latent heterogeneity of a
causally important trait, the solution would seem to be to identify, and then
control for, the latent trait.  ``Homophily'' means simply that individuals
tend to choose neighbors that resemble them; this tendency will be especially
pronounced if pairs of neighbors also have other neighbors in common, since
these pairings will also be driven by homophily. This suggests that homophily,
latent or manifest, will tend to produce a network built primarily of
homogeneous clusters, also called, in this context, ``communities'' or
``modules''.  Inversely, such clusters will tend to consist of nodes with the
same value of the homophilous trait.

The topic of community discovery --- essentially, dividing graphs into
homogeneous, densely inter-connected clusters of nodes, with minimal connection
between clusters --- has been thoroughly explored in the recent literature
(explicitly in \citet{Girvan-MEJN-community-structure,
  MEJN-Girvan-community-structure, Bickel-Chen-on-modularity,
  Porter-Onnela-Mucha-communities, Fortunato-community-detection}, implicitly
in much smaller clusters in \citet{Elwert-Christakis-wives-and-ex-wives}). A
natural idea would be to first establish the existence of these clusters, to
note the memberships of each individual in the chosen model, call this estimate
$\hat{C}_i$, and to control for $\hat{C}_i$ when looking for evidence of
contagion or influence.

By the arguments we have presented so far, such control-by-clustering will
generally be unable to eliminate the confounding\footnote{The exception will be
  if $\hat{C}_i$ was a predictively sufficient statistic, which in this case
  would mean that the realized graph $A$ provided enough information to render
  the true community memberships of all nodes conditionally independent of
  their observed behaviors.  Then we would effectively move from the situation
  of Figure \ref{fig:latent-homophily} to that of Figure
  \ref{fig:observable-control}$b$, with $\hat{C}_i$ in the role of $Z_i$.
  Determining the class of network models for which such ``screening off''
  holds is the subject of on-going work.}.  However, in conjunction with the
bounds approach mentioned above, conditioning on estimated community
memberships might still noticeably reduce the confounding.  On the other hand,
misspecification of the block structure may make the problem worse --- consider
the cases where the generating mechanism may be a mixed-membership block model
\citep{airoldi2008mmsb} or ``role'' model \citep{Reichardt-White-role-models}
but communities are ``discovered'' assuming a simple modular network
structure. Estimating the damage due to misspecification in this case is a goal
of future research.

\section{Conclusion: Towards Responsible Just-So Story-Telling}
\label{sec:conclusion}

We have seen that when there is latent homophily, contagion effects are
unidentifiable, and even the presence of contagion cannot be distinguished
observationally from a causal effect of the homophilous trait.  Conversely,
when contagion and homophily both exist, choices can be predicted from the
homophilous trait, and so the effects of such traits on socially influenced
variables is again observationally unidentifiable.  These results raise
barriers to many inferences social scientists would like to make.  The barriers
can be breached by assuming enough about the causal architecture of the process
in question, though then the inferences stand or fall with those architectural
assumptions; perhaps the bounding approach can squeeze an opening through them
as well.  Beyond these technical qualifications, what is the larger moral for
social science?

Accounts of social contagion are fundamentally causal accounts, pointing to one
of a number of mechanisms --- imitation, persuasion, etc. --- by which a belief
or behavior spreads through a population.  Similarity among individuals is
explained by their belonging to common networks; differences by differences in
their networks.  This parallels the other great project of social science,
which is to explain differences in cultural choices by location within the
social structure, or, at a broader scale, by differences between social
structures \citep{Boudon-ideology, Berger-culture, Lieberson-matter-of-taste}.
The accounts that have connected social structure to behavior have typically
been adaptationist or functionalist: the content or meaning of cultural choices
serves the choosers' interests, or their classes' interests, or (far more
nebulously) the interests of the system, or reflects their experiences in life,
or rationalizes their positions in life, and so forth.  At the very least,
these are causal accounts: if social structure or social positions were
different, the {\em content} of the choices would be different.  Far more
commonly, they really are adaptationist accounts: choices {\em fit} to the
objective circumstances.  They accordingly follow the familiar pattern of the
``Just-So'' story \citep{Kipling-just-so}, with all their familiar problems.
It would be intellectually irresponsible to accept such accounts, with their
strong causal claims, without careful checking; but also irresponsible to
simply dismiss them out of hand.

The example of biology suggests that a powerful way of doing such tests is to
use ``neutral models'' \citep{Harvey-Pagel, Gillespie-pop-gen}, which biologists
use to test claims that features of organisms are evolutionary adaptations; we
note the similarity with the ``null hypothesis'' in general statistical
hypothesis testing.  A neutral evolutionary model should include all the
relevant features of the evolutionary process {\em except} adaptive forces
(such as natural or sexual selection).  The expected behavior of the system is
then calculated under the neutral model (namely, the distribution of expected
outcomes); if the data depart significantly from the predictions of the neutral
model, this is taken as evidence of adaptation.  Said another way, the neutral
model as a whole is used as the null hypothesis, not just a generic regression
model with some coefficients set to zero.  For instance, a model might include
mutation and genetic recombination, but assume all organisms are equal likely
to be parents of the next generation; all have equal fitness.  Gene frequencies
will change in such a model because of random fluctuations; some organisms
become parents and have differing numbers of offspring. Indeed, we expect some
genetic variants to go to fixation (to become universal) in the population, and
others to disappear entirely through the effects of repeated
sampling.\footnote{Superficially, this looks very much like the effects of
  selection, even though the statistical properties of fixation via sampling
  and fixation via selection are quite different; in particular, fixation via
  selection is much faster.} We are not aware of any studies in the sociology
of culture or related fields employing formal neutral models; however,
something similar to this is implicit in the arguments of
\citet{Lieberson-matter-of-taste},\footnote{\citet{Lieberson-Lynn-wrong-branch},
  while offering evolutionary biology as a methodological model for social
  science, curiously does not mention the issue of neutral models.} and some
other strands of recent work on ``endogenous explanations of culture''
\citep{Kaufman-endogenous}.

The point is not that accounts of causation and adaptation in social phenomena
must be rejected; it is that they must be subjected to critical scrutiny, and
that comparison to neutral models is a particularly useful form of critique.
Our toy models produce the kind of phenomena which theories of contagion, or of
adaptation and reflection, set out to explain.  (It is only too easy to imagine
crafting a historical narrative for Figure \ref{fig:network-evolutions},
explaining the deep forces that impelled the east to become red.)  The best way
forward for advocates of those theories may in fact be to craft better, more
compelling neutral models than ours, and show that even these cannot account
for the data.  Thus they will support their theories not only by plausible
just-so stories, but by compelling evidence.

\subsection*{Acknowledgments}

We thank Edo Airoldi, Tanmoy Bhattacharya, Joe Blitzstein, Aaron Clauset, Felix
Elwert, Stephen Fienberg, Clark Glymour, Justin Gross, Matthew Jackson, Brian
Karrer, Kristina Klinkner, David Lazer, Martina Morris, Mark Newman, J{\"o}rg
Reichardt, Martin Rosvall, Richard Scheines, Peter Spirtes, Douglas R. White
and Jon Wilkins for valuable discussions, and the anonymous referees for useful
suggestions.  Preliminary versions of this work were presented at the Santa Fe
Institute workshop on ``Statistical Inference for Complex Networks'', MERSIH 2,
the NIPS workshop ``Analyzing Networks and Learning with Graphs'', and the
Carnegie Mellon seminar on relational learning; we thank the organizers for
their generous hospitality and the participants for their feedback.  Our work
was supported in part by NIH grant 2 R01 NS047493 (CRS) and DARPA grant
21845-1-1130102 (ACT).

\bibliographystyle{crs}
\bibliography{locusts}

\end{document}